\documentclass[12pt,preprint]{aastex}
\usepackage{mathptmx}
\usepackage{natbib}
\usepackage{epsf}
\usepackage{colordvi}

\shorttitle{Different magnetic flux distribution}
\shortauthors{Zhang et al.}

\begin{document}

\title{COMPARISON OF MAGNETIC FLUX DISTRIBUTION \\
BETWEEN A CORONAL HOLE AND A QUIET REGION}

\author{JUN ZHANG\altaffilmark{1, 2}, JUN MA\altaffilmark{2}
and HAIMIN WANG \altaffilmark{2}}
%Vasyl Yurchyshyn\altaffilmark{2}}
%N. A. Schwadron\altaffilmark{3}\\
%AND L. A. Fisk\altaffilmark{4}}

%\email{zjun@ourstar.bao.ac.cn}

\altaffiltext{1}{National Astronomical Observatories, Chinese
Academy of Sciences, Beijing 100012, China;
zjun@ourstar.bao.ac.cn} \altaffiltext{2}{Big Bear Solar
Observatory, New Jersey Institute of Technology, Big Bear City, CA
92314; haimin@flare.njit.edu, junma@bbso.njit.edu}
%vayur@bbso.njit.edu}
%\altaffiltext{3}{Boston University, Dept. of
%Astronomy, 725 Commonwealth Avenue, Boston, Massachusetts 02215;
%nathanas@bu.edu}
%\altaffiltext{4}{Department of Atmospheric,
%Oceanic, and Space Sciences, University of Michigan, 2455 Hayward
%Street, Ann Arbor, MI 48109-2143; lafisk@umich.edu}

\begin{abstract}

Employing Big Bear Solar Observatory (BBSO) deep magnetograms and
H${\alpha}$ images in a quiet region and a coronal hole, observed
on September 14 and 16, 2004, respectively, we have explored the
magnetic flux emergence, disappearance and distribution in the two
regions. The following results are obtained: (1) The evolution of
magnetic flux in the quiet region is much faster than that in the
coronal hole, as the flux appeared in the form of ephemeral
regions in the quiet region is 4.3 times as large as that in the
coronal hole, and the flux disappeared in the form of flux
cancellation, 2.9 times as fast as in the coronal hole. (2) More
magnetic elements with opposite polarities in the quiet region are
connected by arch filaments, estimating from magnetograms and
H${\alpha}$ images. (3) We measured the magnetic flux of about
1000 magnetic elements in each observing region. The flux
distribution of network and intranetwork (IN) elements is similar
in both polarities in the quiet region. For network fields in the
coronal hole, the number of negative elements is much more than
that of positive elements. However for the IN fields, the number
of positive elements is much more than that of negative elements.
(4) In the coronal hole, the fraction of negative flux change
obviously with different threshold flux density. 73\% of the
magnetic fields with flux density larger than 2 Gauss is negative
polarity, and 95\% of the magnetic fields is negative, if we only
measure the fields with their flux density larger than 20 Gauss.
Our results display that in a coronal hole, stronger fields is
occupied by one predominant polarity; however the majority of
weaker fields, occupied by the other polarity.

\end{abstract}


\section{INTRODUCTION}

Small-scale magnetic elements on the Sun have been under intensive
investigation for many years. They occur predominantly in the
quiet sun region, especially along the borders of the
supergranular cells where the chromospheric network can be
visualized in the filtergrams of chromospheric lines(\Red{need
line examples}). Based on their locations and morphology(\Red{am I
right here?}), small-scale magnetic fields in the quiet sun region
can be classified in three categories, namely, network
\citep{lei62}, intranetwork (IN) \citep{liv75} and ephemeral
regions \citep{har73}.

Based on the polarimetry results using the fourier transform
spectrometer \Red{(reference)} at Kitt Peak, Arizona, together
with sophisticated modelling, the Z\"{u}rich group and its
collaborators were able to establish hydrodynamic pictures for
small scale magnetic flux tubes in plages and network of the quiet
Sun. Extensive reviews on these results can be found in
\citet{sol93} and \citet{ste94}. The first Stokes-V measurement
using line ratio technique \Red{(reference)} was made by
\citet{kel94}, who placed an upper limit on the intrinsic strength
of intranetwork (IN) fields to be 1000 gauss or 500 gauss with
68\% probability. Also using infrared spectro-polarimetry,
\citet{lin95} found that the typical field strength of IN fields
are around 500 gauss. \citet{san96} suggested that the IN fields
are highly irregular over optically thin scales \Red{(layers?)}.
\citet{kne96} presented an image in which small-scale magnetic
elements possess substructure and are dynamical, with gas flows
and magnetic field strength varying in space and time.
\citet{meu98} obtained the fraction of magnetic flux in a weak
field form, i.e. with magnitude lower than 1,000 Gauss intrinsic
strength in the quiet Sun. However, \citet{soc04} presented an
evidence of strong ($\sim$1,700~Gauss) and weak ($<$500 Gauss)
fields coexisting within the resolution element at both network
and IN regions, and there was a larger fractional area of weak
fields in the convective upflows than in the downflows.

\Red{(I guess more specific contents of each paper referred here
need to be mentioned.)}In parallel, by means of video magnetograph
observations, especially based on time sequences of deep
magnetograms obtained at Big Bear Solar Observatory (BBSO), much
progress was made in network and IN morphology dynamics and some
quantitative aspects, such as flux distribution of quiet-Sun
magnetic elements \citep{wan95}; mean horizontal velocity fields
of IN and network fields by using the local correlation tracking
technique \citep{wan96}; lifetime of IN elements \citep{zha98a};
motion patterns and evolution of IN and network magnetic fields
\citep{zha98b, zha98c}.

On the other hand, coronal holes are cool, low density regions,
which can be observed at both low latitudes region and polar
region of the Sun \citep{chi99}. They were first observed on X-ray
plates by \citet{und67}, on EUV line spectroheliograms by
\citet{ree70} and in white light by \citet{alt72}. The magnetic
fields within a coronal hole region are usually dominated by one
polarity, and thus the field lines in the upper atmosphere are
open to the interplanetary region \citep{boh77}, generating
high-speed solar wind that can lead to geomagnetic storms
\citep{kri73}. According to the location and lifetime, there are
three categories of coronal holes: polar, non-polar and transient
coronal holes. Polar coronal holes have long lifetimes (about 8
years)\Red{(reference)}. Non-polar coronal holes are usually
associated with remnants of active regions and may persist for
many solar rotations. Transient coronal holes are associated with
eruptive events, such as filament eruptions and coronal mass
ejections (CMEs) and have lifetimes of several days \citep{har02}.
Also, low-latitude coronal holes may show quasi-rigid rotation and
it has been suggested that magnetic reconnection must occur
continuously at the boundary in order to maintain the integrity of
coronal hole \citep{kah02}.

In order to answer some of the key questions in solar and stellar
physics, such as the coronal heating and solar wind acceleration,
we need to study and understand the small-scale magnetic activity
both in coronal holes and in quiet regions. In this paper, we
focus on two topics: a) the dipole flux emergence and flux
disappearance in a coronal hole and a quiet region; b) the
distribution of magnetic flux with both polarities in the two
observing regions, using BBSO long-integration magnetograms. Base
on these discussion, we will finally propose a physical model for
the characteristics of the magnetic fields in these two regions.

\section{OBSERVATIONS AND ANALYSIS}
A sequence of collaborative observations were carried out from
13th to 18th in September 2004 by BBSO team and the Transition
Region and Coronal Explorer (TRACE) team. We selected the
observation data obtained on 14th and 16th for this study due to
the completeness and the quality of the data. The observation area
on 14th was a quiet region centered at N9\arcdeg W13\arcdeg, and
on 16th, was a coronal hole at N31\arcdeg E14\arcdeg. H${\alpha}$
line center and Ca\,{\sc ii} K line images were taken with two
Kodak CCD cameras mounted on the 25~cm refractor with 90~s
cadence. Images at H${\alpha}{\pm}$0.60 {\AA} were taken with an
Orbiting Solar Laboratory (OSL) CCD camera mounted on the 65 cm
telescope with a 30 s cadence, which have FOV
210$''$${\times}$210$''$, and an image scale of 0\arcsec.4
pixel$^{-1}$. The magnetogram was obtained using the digital
vector magnetograph (DVMG) system mounted on the 25~cm refractor
with a cadence of 90~s, FOV 300\arcsec $\times$ 300\arcsec, and an
image scale of 0\arcsec.60 pixel$^{-1}$. The DVMG system uses a
Zeiss filter, and a 12-bit digital camera. The averaged noise
level of these magnetogram is 2~Gauss. In Figure.~1, the
magnetogram of the quiet region (upper panel) and the coronal hole
(lower panel) are shown together for comparison.

The high-resolution UV observation was obtained from TRACE
satellite \citep{han99}, which are used in this paper as a
reference for identifying fine structures and the dynamics of the
corona and transition region. The 1600~\AA\, images used in this
study have spatial resolution of 1\arcsec, temporal resolution of
40~s, and field-of-view of 250\arcsec ${\times}$250\arcsec.

\Red{I think there should be more text regarding the analysis
process. How did you process the data? How did you identify the
features? etc.}

\section{RESULTS}

\subsection{MAGNETIC FLUX EMERGENCE AND DISAPPEARANCE}

%\citet{wan92} studied the properties of ephemeral regions (ERs) in
%a coronal hole.
In the co-spatial field-of-view (about 200${\times}$200 square
arcsec) among BBSO magnetograms, H${\alpha}$ filtergrams, and
TRACE 1600 {\AA} images, We identified 30 pairs of ephemeral
regions (ERs) in the quiet region on September 14, 2004 during the
7-hour observations. The mean flux of an ER is
8.1${\times}$10$^{18}$ Mx. 7 of the 30 ERs are connected by arch
filaments, seeing from H${\alpha}$ images. Figure 2 shows the
evolution of an ER (see the two arrows on the magnetogram at 20:36
UT). H${\alpha}+$0.60 {\AA} observations display that the two
elements of the ER always undergo downflows (Doppler blueshifts,
black patches indicated by two arrows on the H${\alpha}+$0.60
{\AA} image at 20:35 UT). Just as the appearance of the ER, UV
bright points (the arrows on the UV image at 20:41 UT) also appear
at the feet of the two magnetic elements, seen from TRACE 1600
{\AA} images. As the ER growing, it is more clearly that the two
elements belonging to the ER are connected by an arch filament
(denoted by dotted lines in two H${\alpha}$ filtergrams at 22:49
and 22:50 UT) and the UV bright points become larger and larger,
then each point splits into several ones(e.g., the arrows in the
bottom panel).

For the coronal hole, although the observational time and the
field-of-view are the same as that of the quiet region, we can
only track 17 pairs of ERs, and one of the 17 ERs is connected by
an arch filament. The mean flux of an ER is 3.4${\times}$10$^{18}$
Mx, much smaller than that in the quiet region. Figure 3 displays
the flux versus flux density of all the ERs in the quiet region
and the coronal hole. It shows that there is a similar
flux and flux density distributions of the ERs in the two
regions. However in the coronal hole, the number of the ERs in a
higher flux density and a larger flux range is much smaller than
that in the quiet region, and the total flux of all the 17 ERs is
5.8${\times}$10$^{19}$ Mx, only the 23\% as large as the flux
(2.5${\times}$10$^{20}$ Mx) of the 30 ERs in the quiet region.
This implies that the evolution of magnetic flux in the coronal
hole is much slower than that in the quiet region. We find from
Figure 3 that only if the flux is larger than 5.0${\times}$10$^{18}$
Mx and flux density higher than 20 Gauss then the ERs are observed
to be connected by arch filament systems. Considering ERs as
closed magnetic loops, we can estimate the flux density of these
loops ranges from 6 to 40 Gauss in the quiet region, and from 6 to
22 Gauss in the coronal hole.

Another parameter which can also represent the magnetic evolution
rate is magnetic flux disappearance. The disappearance is mainly
due to the cancellation of opposite polarity elements. If an
magnetic element meets with another of opposite polarity, they can
cancel each other. Under this condition, the total flux decreases.
In the quiet region, about 1.7${\times}$10$^{20}$ Mx
(2.1${\times}$10$^{20}$ Mx) positive flux (negative flux)
disappear by the cancellation in the 7-hour observations. In the
coronal hole, the amount of the disappeared flux is about
3.5${\times}$10$^{19}$ Mx (9.6${\times}$10$^{19}$ Mx) for positive
flux (negative flux), one third as large as that in the quiet
region.

\subsection{MAGNETIC FLUX DISTRIBUTION}

It is always suggested that coronal holes lie within a unipolar
magnetic region. However the solar magnetic field is never
strictly unipolar. We have no knowledge about the IN flux
distribution in coronal holes, although \citet{wan95} presented
flux distribution of IN and network magnetic elements in a quiet
region. Here we study the flux distribution in both polarities in
the coronal hole, and compare with the quiet region. Three
criteria are used to separate the IN elements from the network
elements \citep{wan95}. For each region, about 1000 magnetic
elements are identified and measured. The statistics are listed in
Table 1. Figure 4 presents flux distributions of all the measured
magnetic elements. The bin size for IN data points is
5.0${\times}$10$^{16}$ Mx, and 5.0${\times}$10$^{17}$ Mx for
network. In the quiet region, the flux distribution of IN and
network elements is similar in both polarities. For network fields
in the coronal hole, there is no positive elements in the range
where magnetic flux larger than 4.0${\times}$10$^{18}$ Mx. However
for the IN fields, the number of positive elements is much larger
than that of negative elements. In other words,
in this coronal hole the stronger field is occupied by
one predominant polarity; and the majority of weaker fields,
occupied by the opposite polarity.

Now we measure the magnetic flux in both polarities instead of the
individual elements in the two regions. Figure 5 shows the
variation of magnetic flux versus threshold flux density in the
field-of-view of the magnetograms in Figure 1. As the threshold
flux density increases from 2 Gauss (near noise level) to 30
Gauss, the flux in both polarities decreases homologically in the
quiet region (see the upper panel of Figure 5). In the coronal
hole, the positive flux decreases more quickly than the negative
flux with increasing threshold flux density, as shown in the
middle panel of Figure 5. The bottom panel presents the fraction
of the negative flux to the total (negative plus positive) flux.
In the coronal hole, 73\% of the magnetic fields with flux density
larger than 2 Gauss is negative, and 95\% of the magnetic fields
is negative, if we only measure the fields with their flux density
larger than 20 Gauss, the noise level of a typical magnetogram.
Comparing to
the coronal hole, we conclude that the fraction
in the quiet region changes more gently. This result further
confirms the conclusion that in the coronal hole, stronger field is
occupied by one predominant polarity, and weaker fields, occupied
by the opposite polarity with a large margin.

\section{DISCUSSION}

In this paper, we find that the dipole flux emergence rate in the
quiet region exceeds that in the coronal hole by thrice. A similar
result was presented by \citet{abr06} who used the MDI data with
the detection limit for the magnetic flux density being 17 Gauss,
and only looked at the much larger ERs. This result is consistent
with the model of \citet{fis05}. In his model, the regions where
the rate of emergence of new flux is a local minimum, open flux
accumulates to form coronal holes. The flux disappearance rate in
the quiet region is also more than twice larger than that in the
coronal hole. Therefore, the transformation rate from magnetic
energy to heat and kinetic energy is lower in the coronal hole.

H${\alpha}$ observations show that many H${\alpha}$ threads
connect dipolar magnetic elements in the quiet region. However in
the coronal hole, almost all the H${\alpha}$ threads appear as
jets in shape. This means that near the atmospheric level where
H${\alpha}$ line forms, the magnetic fields in coronal hole are
almost opened. Magnetic field measurement indicates that in the quiet
region the flux distribution of the IN fields is similar to that of the
network fields, and the opposite polarity fluxes are basically balanced.
In the coronal hole, the number of negative network elements is
much more dominant than that of positive elements. However for the IN
fields, the number of positive elements is much more than that of
negative elements. We search some other high resolution magnetic
fields data to check the reliability of the results shown in
Figure 5. The BBSO deep magnetograms of June 4, 1992 provided
unprecedented observations for a quiet sun. The variation of
magnetic flux versus threshold flux density is similar as that in
the quiet region observed on Sep. 14, 2004. In other two coronal
hole regions which located near the solar disk, observed on
September 17, 2004 and October 11, 2005, the magnetic flux
variation is also followed the tendency as shown in the middle
panel of Figure 5. This implies that the different magnetic flux
distribution between the coronal hole and the quiet region is
somehow common, not just a random case.

The observed scenario for the magnetic structures in the two
regions can be schematically shown in Figure 6. Dotted lines
represent the magnetic field lines which have no H${\alpha}$
counterparts. The loops in the coronal hole are on average flatter
than in the quiet region. High and long closed loops are extremely
rare, whereas short and low-lying loops are almost as abundant in
coronal holes as in the quiet Sun \citep{wie04}. The observations
may hint a physical picture that some IN flux, preferentially
close to the cell boundary, may be topologically connected to the
network field \citep{zha99}. It is necessary to point out that IN
flux of the same sign of the surrounding network is more likely to
be counted as network flux. Somehow IN flux of the same sign as
the surrounding network is destroyed more rapidly either by
merging with the network or diffusing. Furthermore, IN flux of the
opposite polarity rises in more concentrated form and so it is
easier to be detected.

\citet{tu05} established that the fast solar wind starts flowing
out of a corona hole at heights above the photosphere between
5,000 and 20,000 kilometers in magnetic funnels. Our work shows
that in the coronal hole on Sep. 16, 2004, most of closed magnetic
loops are lower than 5,000 kilometers, and open magnetic field
with one polarity fill the space above 5,000 kilometers, where
fast solar wind originates from. Flux cancellation, or a
lower magnetic reconnection in the photosphere and lower
chromosphere may only take place below the atmospheric level of
5,000 kilometers, not in the location where the fast solar wind
starts. This implies that the release of magnetic energy and the
origin of fast solar wind happen at different atmospheric level.

\acknowledgments

The authors are indebted to the TRACE and BBSO teams for providing
the wonderful data. We thank Dr. Yohei Yamauchi for organizing the observing
campaign. The work is supported by the National Natural
Science Foundations of China (G10573025 and G10233050), the State
Ministry of Science and Technology of China under Grant of
G200078404, and two US NASA grants (NAG5-12780 and NNG0-4GG21G)

\clearpage
\begin{table}{}
\begin{center}
%\caption{
TABLE 1 \\
Flux distribution in the two regions in both polarities \\
%\label{tbl-1}}
\vspace{0.5cm}
\begin{tabular}{lllll}
\tableline\tableline \label{tbl-1}
 & Network & & IN &                               \\
The quiet sun & Positive & Negative & Positive & Negative  \\
\tableline
Number & 102 & 116 & 376 & 383                       \\
Total flux( 10$^{20}$ Mx)& 4.1 & $-$6.4 & 2.2 & $-$2.8      \\
Flux imbalance & & $-$0.22 &  & $-0.12$     \\
Mean flux (10$^{18}$ Mx)& 4.02 & $-$5.52 & 0.59 & $-$0.73     \\
Flux density (Gauss)    & 21.3 & $-$25.8 & 4.1 & $-$4.6    \\

\tableline
The coronal hole &  &  &  &                        \\
\tableline
Number & 59 & 153 & 540 & 342                      \\
Total flux (10$^{20}$ Mx)& 1.1 & $-$9.8 & 2.7 & $-$1.6      \\
Flux imbalance & & $-$0.80 &  0.26  &    \\
Mean flux (10$^{18}$ Mx)& 1.86 & $-$6.41 & 0.50 & $-$0.47        \\
Flux density (Gauss)    & 15.2 & $-$29.5 & 4.2 & $-$4.3     \\
\tableline
\end{tabular}
\end{center}
\end{table}

\begin{figure}
\figurenum{1}
\epsscale{0.60}
%\plotone{f1.eps}
\figcaption[f1.eps]{BBSO magnetograms in a quiet region (upper)
and a coronal hole (lower). The field of view is 200${\times}$200
square arcsec. A box in the magnetogram of the quiet sun outlines
a region of ephemeral flux (see Fig. 2). \label{fig1}}
\end{figure}

\begin{figure}
\figurenum{2}
\epsscale{0.90}
%\plotone{f2.eps}
\figcaption[f2.eps]{A pair of ephemeral region in the quiet region
on September 14, 2004. From left colume to right column: BBSO
magnetograms, H${\alpha}$$-$0.6 {\AA} images, H${\alpha}$+0.6
{\AA} images, H${\alpha}$ Dopplergrams, and {\it TRACE} 1600 {\AA}
images. The field of view is about 30${\times}$30 square arcsec.
Arrows and dotted lines are described in the text. \label{fig2}}
\end{figure}

\begin{figure}
\figurenum{3}
\epsscale{0.90}
%\plotone{f3.eps}
\figcaption[f3.eps]{{\it Upper:} Flux vs flux density of 30 pairs
of ERs in the quiet region. {\it Lower:} The same as upper panel
but for 17 pairs of ERs in the coronal hole. Vertical dotted lines
show the flux density of 20 Gauss, and horizontal lines, flux of
5${\times}$10$^{18}$ Mx. Heavy symbols represent these ephemeral
regions that are connected by arch filament systems. \label{fig3}}
\end{figure}

\begin{figure}
\figurenum{4}
\epsscale{1.00}
%\plotone{f4.eps}
\figcaption[f4.eps]{{\it Upper:} Flux distributions of positive
and negative elements in the quiet region (upper) and the coronal
hole (lower). Vertical dotted lines represent the magnetic flux of
10$^{18}$ Mx which separates the IN elements from the network
elements. \label{fig4}}
\end{figure}

\begin{figure}
%\figurenum{5} \epsscale{0.60} \plotone{f5.eps}
\figcaption[f5.eps]{Variation of magnetic flux vs threshold flux
density in the field-of-view of magnetograms in Fig. 1 in the
quiet region ({\it upper}) and the coronal hole ({\it middle}).
The lower panel displays the flux fraction of the negative flux
relative to the total flux. \label{fig5}}
\end{figure}

\begin{figure}
\figurenum{6}
\epsscale{1.00}
%\plotone{f6.eps}
%\vspace{-6.0cm}
\figcaption[f6.eps]{Schematic of the magnetic structure of two
regions. {\it Upper:} The quiet sun. {\it Lower:} The coronal
hole. Solid lines indicate the magnetic field lines which have
H${\alpha}$ counterparts, e.g. the closed lines versus arch
filaments, and opened lines, macrospicules. The dotted lines show
field lines which have no H${\alpha}$ counterparts. \label{fig6}}
\end{figure}

\clearpage


\begin{thebibliography}{}
\bibitem[Abramenko, Fisk, \& Yurchyshyn(2006)]{abr06} Abramenko,
V. I., Fisk, L. A., \& Yurchyshyn, V. B. 2006, \apjl, accepted
\bibitem[Altshuler \& Perry(1972)]{alt72}
Altshuler, M. D., \& Perry, R. M. 1972, \solphys, 23, 410
\bibitem[Bohlin(1977)]{boh77} Bohlin, J. D. 1977, \solphys, 51, 377
\bibitem[Chiuderi Drago et al.(1999)]{chi99} Chiuderi Drago, F.,
Landi, E., Fludra, A., \& Kerdraon, A. 1999, \aap, 348,
261
\bibitem[Fisk(2005)]{fis05} Fisk, L. A. 2005, \apj, 626, 563
\bibitem[Handy et al.(1999)]{han99} Handy, B. N., et al.
1999, \solphys, 187, 229
\bibitem[Harvey \& Martin(1973)]{har73}
Harvey, K. L., \& Martin, S. F. 1973, \solphys, 32, 389
\bibitem[Harvey \& Recely(2002)]{har02} Harvey, K. L., \& Recely, F.
2002, \solphys, 211, 31
\bibitem[Kahler \& Hudson(2002)]{kah02}
Kahler, S. W., \& Hudson, H. S. 2002, \apj, 574, 467
\bibitem[Keller et al.(1994)]{kel94} Keller,
C. J., Deubner, F. L., Egger, U., Fleck, B., \& Povel, H. P. 1994,
\aap, 286, 626
\bibitem[Kneer \& Stolpe(1996)]{kne96} Kneer, F., \& Stolpe, F. 1996,
\solphys, 164, 303
\bibitem[Krieger \& Timothy(1973)]{kri73} Krieger, A. S., \& Timothy, A. F.
1973, \solphys, 29, 505
\bibitem[Leighton, Noyes, \& Simon(1962)]{lei62} Leighton, R. B.,
Noyes, R. W., \& Simon, G. W. 1962, \apj, 135, 474
\bibitem[Lin(1995)]{lin95} Lin, H. 1995, \apj, 446, 421
%\bibitem[Lites(2002)]{lit02}Lites, B. W. 2002, \apj , 573, 431
\bibitem[Livingston \& Harvey(1975)]{liv75} Livingston, W. C., \&
Harvey, J. 1975, \baas, 7, 346
\bibitem[Meunier, Solanki, \& Livingston(1998)]{meu98} Meunier, N.,
Solanki, S. K., \& Livingston, W. C. 1998, \aap, 331, 771
\bibitem[Reeves \& Parkinson(1970)]{ree70} Reeves, E. M., \& Parkinson, W. H.
1970, \apjs, 21, 1
\bibitem[S\'{a}nchez Almeida et al.(1996)]{san96} S\'{a}nchez Almeida, J.,
Landi Degl'innocenti, E., Mart\'{\i}nez Pillet, V., \& Lites, B.
W. 1996, \apj, 466, 537
\bibitem[Socas-Navarro \& Lites(2004)]{soc04} Socas-Navarro, H., \& Lites, B. W.
2004, \apj , 616, 587
\bibitem[Solanki(1993)]{sol93} Solanki, S.
K. 1993, \ssr, 63, 1
\bibitem[Stenflo(1994)]{ste94} Stenflo, J. O. 1994, Solar Magnetic Fields $-$
Polarized Radiation Diagnostics, Kluwer, Dordrecht
\bibitem[Tu et al.(2005)]{tu05} Tu, C., Zhou, C., Marsch, E. et al. 2005,
Science, 308, 519
\bibitem[Underwood \& Muney(1967)]{und67}
Underwood, J. H., \& Muney, W. S. 1967, \solphys, 1, 129
\bibitem[Wang et al.(1996)]{wan96} Wang, H., Tang, F., Zirin, H., Wang, J.
1996, \solphys, 165, 223
\bibitem[Wang, Wang, \& Shi(1992)]{wan92} Wang, J., Wang, H., \& Shi, Z. 1992,
ASP Conference Series, 27, 108
\bibitem[Wang(1995)]{wan95} Wang, J., Wang, H., Tang, F., et al. 1995,
\solphys, 160, 277
\bibitem[Wiegelmann \& Solanki(2004)]{wie04} Wiegelmann, T.,
\& Solanki, S. K. 2004, \solphys,  225, 227
\bibitem[Zhang et al.(1998a)]{zha98a} Zhang, J.,
Lin, G., Wang, J., Wang, H., \& Zirin, H. 1998a, \solphys, 178,
245
\bibitem[Zhang et al.(1998b)]{zha98b} Zhang, J., Wang, J., Wang, H., \&
Zirin, H. 1998b, \aap, 335, 341
\bibitem[Zhang et al.(1998c)]{zha98c} Zhang, J., Lin, G., Wang, J., Wang, H.,
\& Zirin, H. 1998c, \aap, 338, 322
\bibitem[Zhang et al.(1999)]{zha99} Zhang, J., Wang, J., Deng, Y.,
\& Wang, H. 1999, \solphys, 188, 47
\end{thebibliography}
\end{document}